**Dirac points in helically structured 1D photonic crystals**

A.H. Gevorgyan

Far Eastern Federal University, 10 Ajax Bay, Russky Island, Vladivostok 690922,

Russia; e-mail: agevorgyan@ysu.am

**Abstract**

We reported about observation of Dirac points in a helically structured 1D photonic crystals, moreover, both as in the presence of longitudinal magnetic field as its absence. We obtained analytical formulas for Dirac points frequencies and the analytical dispersion relations for wave vectors.

Keywords: Dirac point, magnetically induced transparency, magnetically induced absorption, diffraction, eigen modes, indirect photonic bund gap

In recent years, one of the main directions in physics is the search and study of new types of materials characterized by unique electronic, optical, magnetic or mechanical properties, which are described by fundamentally new principles. For example, a number of solid-state systems were discovered in which the dynamics of electrons is described by the relativistic Dirac equation (which is the wave equation formulated to describe relativistic spin 1/2 particles [1]), instead of the classical Schrödinger equation. In the special case, the mass of relativistic fermions can vanish, leading to the formation of a linear dispersion of the electronic states of the Dirac cone in the sense that the energy $E$ is linearly proportional to the wave vector $k$. Graphene is one of such system. Due to the existence of Dirac cones [2], graphene exhibits intriguing transport properties [3-8], such as Klein tunneling [4], Zitterbewegung [5], anti-localization [6], abnormal quantum Hall effect [7,8], etc. Then Dirac points extended to Photonic Crystals (PCs). In [9,10] pointed out the existence of photonic Dirac cones in 2D PCs and discussed the unidirectional propagation of surface modes caused by the time-reversal symmetry breaking. It was showed that if an external magnetic field is used to break time reversal symmetry, unidirectional and backscattering immune electromagnetic wave propagation, analogous to quantum hall edge states, can be realized. Dirac point appears in the first Brillouin zone corners, where the upper and lower bands touch in a linear fashion as to generate a Dirac cone. At this special point, Maxwell’s equations can be replaced by the massless Dirac equation for relativistic particles. Thus, analogies with the electronic energy band structure of graphene [2] can be sustained. In [11] considered properties of Dirac points for honeycomb-lattice PCs. In [12] proposed the realization of Zitterbewegung effect in PCs by the incidence of a pulse near the Dirac point frequency. Pseudo-diffusive transmission with intensity inversely proportional to the thickness of PC array was also discovered in [13] and experimentally demonstrated in [14]. In [15] light trapping peculiarities in PCs in Dirac points were investigated. It is found that the light trapping mechanism of a Dirac point is different from that of a photonic band gap (PBG). For example, one of the new features of a Dirac mode is that the envelope function of the state does not decay exponentially as might be expected. Instead, it decays algebraically away from its center, thus enabling long-range coupling between resonators and waveguides. In the papers [16-18] the peculiarities of Dirac points in 1D PCs were investigated.

In this paper we analytically investigated the peculiarities of Dirac points in helically structured 1D PCs.

We consider the helically structured 1D PC with dielectric permittivity and magnetic permeability tensors having the forms:

$$\hat{\varepsilon}(z) = \varepsilon_m \begin{pmatrix} 1+\delta\cos2az & \pm\delta\sin2az & 0 \\ \pm\delta\sin2az & 1-\delta\cos2az & 0 \\ 0 & 0 & 1-\delta \end{pmatrix}, \ \hat{\mu}(z) = \hat{I}, \tag{1}$$

where $\varepsilon_m = (\varepsilon_1 + \varepsilon_2)/2$, $\delta = \frac{(\varepsilon_1 - \varepsilon_2)}{(\varepsilon_1 + \varepsilon_2)}$, $\varepsilon_{1,2}$ are the principal values of the local dielectric permittivity tensor, $a = 2\pi/p$, $p$ is the pitch of the helix, axis z directed along helix axis. First, we consider the case when the real and imaginary parts of the principal values of the local dielectric permittivity tensor are constant and do not depend on frequency. As well-known, the dispersion equation in the rotating frame (the *x*' and *y*' axes of which rotate together with the structure; besides, the *x*' axis is oriented along the local optical axis everywhere, the *y*' axis is perpendicular to the *x*' axis and *z*' axis directed along medium's helix axis) when light propagates along the helix axis, has the form [19]:

$$\left(\frac{\omega^2}{c^2}\varepsilon_1 - k_z^2 - a^2\right)\left(\frac{\omega^2}{c^2}\varepsilon_2 - k_z^2 - a^2\right) - 4a^2k_z^2 = 0, \qquad (2)$$

where $k_z$ is the *z* component of the wave vector in the rotating coordinate frame. Fig. 1 shows the dependences of $k_{zm}$ (*m*=1,2,3,4) for the frequency $\omega$. Further, there is a frequency band (with the boundaries $\omega_1 = \frac{ca}{\sqrt{\varepsilon_1}}$ and $\omega_2 = \frac{ca}{\sqrt{\varepsilon_2}}$) where two of the four wave numbers are purely imaginary in the absence of absorption ($\mathrm{Im}k_m \neq 0$ and $\mathrm{Re}k_m = 0$); we call them resonant wave numbers. This is where PBG exists, and this PBG is direct one. Let us now enumerate the eigen solutions of Eq. (2) in the following way: *m* = 1 and 4, correspondingly, for the non-resonance wave vectors (for propagation eigenmodes), and *m* = 2 and 3, correspondingly, for the resonance wave vectors (for evanescent eigen modes).

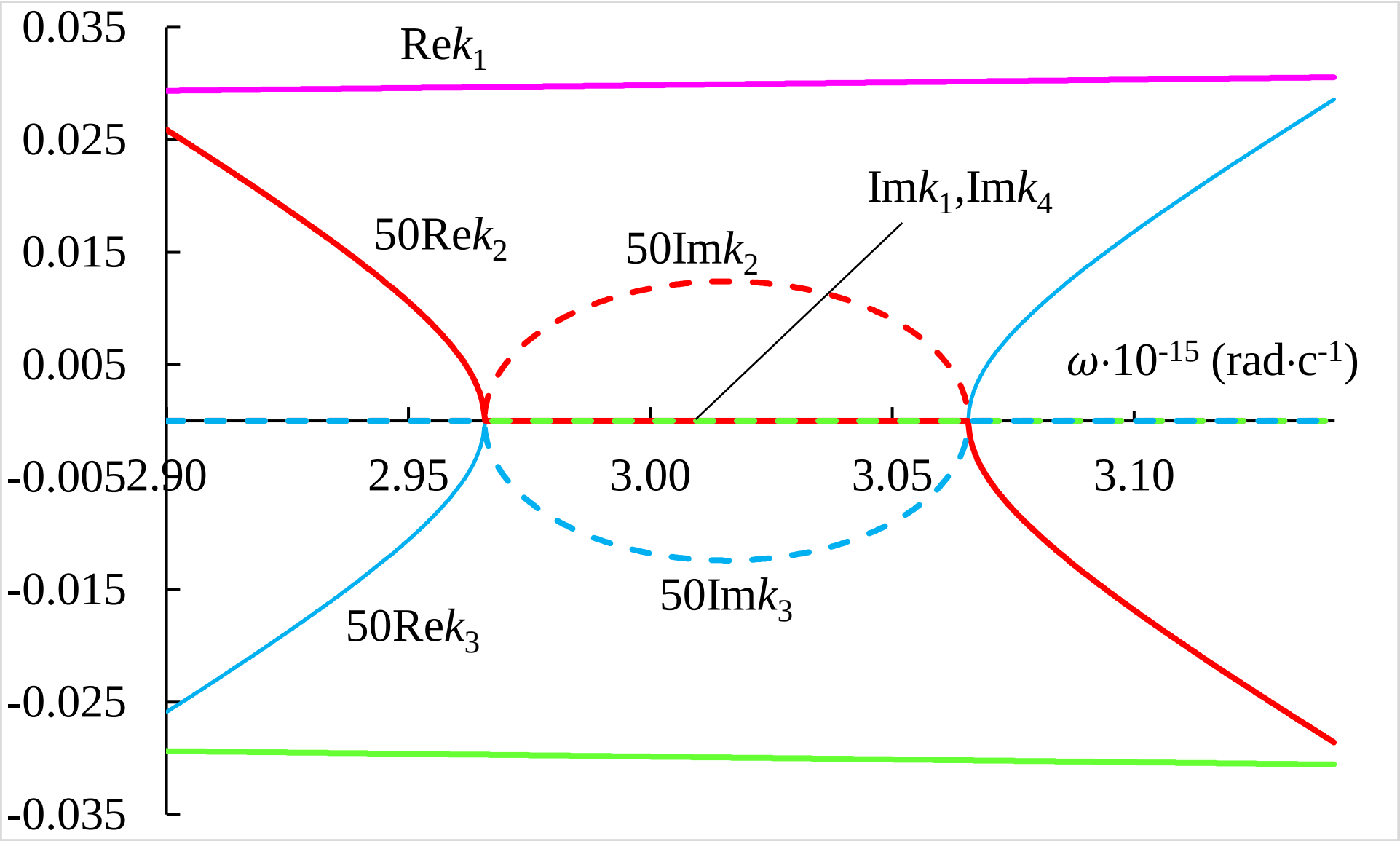

Fig.1. The dependences of $\mathrm{Re}k_{mz}$ and $\mathrm{Im}k_{mz}$ (*m*=1,2,3,4) for the frequency $\omega$. The chiral structured PCs parameters are: $\varepsilon_1 = 2.29 + i0$; $\varepsilon_2 = 2.143 + i0$.; *p*=420 nm.

As in [20], Dirac points in helically structured 1D PCs should be expected when band gap closes, too. Band gap closes when $\varepsilon_1 = \varepsilon_2 = \varepsilon$. In this case Eq. (2) splits into two quadratic equations that have roots:

$$k_{1z} = \frac{\omega}{c}\sqrt{\varepsilon} + a,\ \ k_{2z} = \frac{\omega}{c}\sqrt{\varepsilon} - a,\ k_{3z} = -\frac{\omega}{c}\sqrt{\varepsilon} + a,\ \ k_{4z} = -\frac{\omega}{c}\sqrt{\varepsilon} - a. \qquad (3)$$

Fig. 2 shows the dependences of $k_{zm}$ (*m*=1,2,3,4) for the frequency $\omega$ in this case. Thus, we have the intersection of the wave vector curves and the linear law of the wave vector dependence on the frequency, i.e. here we have a Dirac point. Equating $k_{2z}$ to $k_{3z}$ we obtain an expression for the frequency of the Dirac point: $\omega_D = \frac{ca}{\sqrt{\varepsilon}}$. At the presence of anisotropic absorption ($\mathrm{Re}\varepsilon_1$=$\mathrm{Re}\varepsilon_2$ but $\mathrm{Im}\varepsilon_1 \neq \mathrm{Im}\varepsilon_2$) here is formed new type PBG. A detailed study of this particular case is presented in the works [21-23]. In particular, in [22,23] it was investigated the light trapping peculiarities in this situation, and it was shown that as in [15] the envelope function of the total field on the frequency $\omega_D$ does not decay exponentially with $z$, as might be expected, but algebraically. Moreover, on this frequency the transmission spectrum has a peak, too, as the reflection one.
Further, for $\varepsilon_1 \neq \varepsilon_2$ from (3) we will have

$$k_{1,4z} = \pm\sqrt{\frac{\omega^2}{c^2}\varepsilon_m + a^2 + \gamma},\ k_{2,3z} = \pm\sqrt{\frac{\omega^2}{c^2}\varepsilon_m + a^2 - \gamma}, \quad (4)$$

where $\gamma = \sqrt{\left(\frac{\omega^2}{c^2}\Delta\right)^2 + 4a^2\frac{\omega^2}{c^2}\varepsilon_m}$, $\Delta = (\varepsilon_1 - \varepsilon_2)/2$.
Equating $k_{2z}$ to $k_{3z}$, in this case, we obtain expressions for PBG boundaries $\omega_1 = \frac{ca}{\sqrt{\varepsilon_1}}$ and $\omega_2 = \frac{ca}{\sqrt{\varepsilon_2}}$. In the interval from $\omega_1$ to $\omega_2$ we have $\mathrm{Re}k_{2z} = \mathrm{Re}k_{3z} = 0$, and $\mathrm{Im}k_{2z} = -\mathrm{Im}k_{3z}$ (see, Fig.1). And outside of this interval we have not the linear law of the wave vector dependence on the frequency. Therefore $\omega_1$ and $\omega_2$ are not Dirac points frequencies.

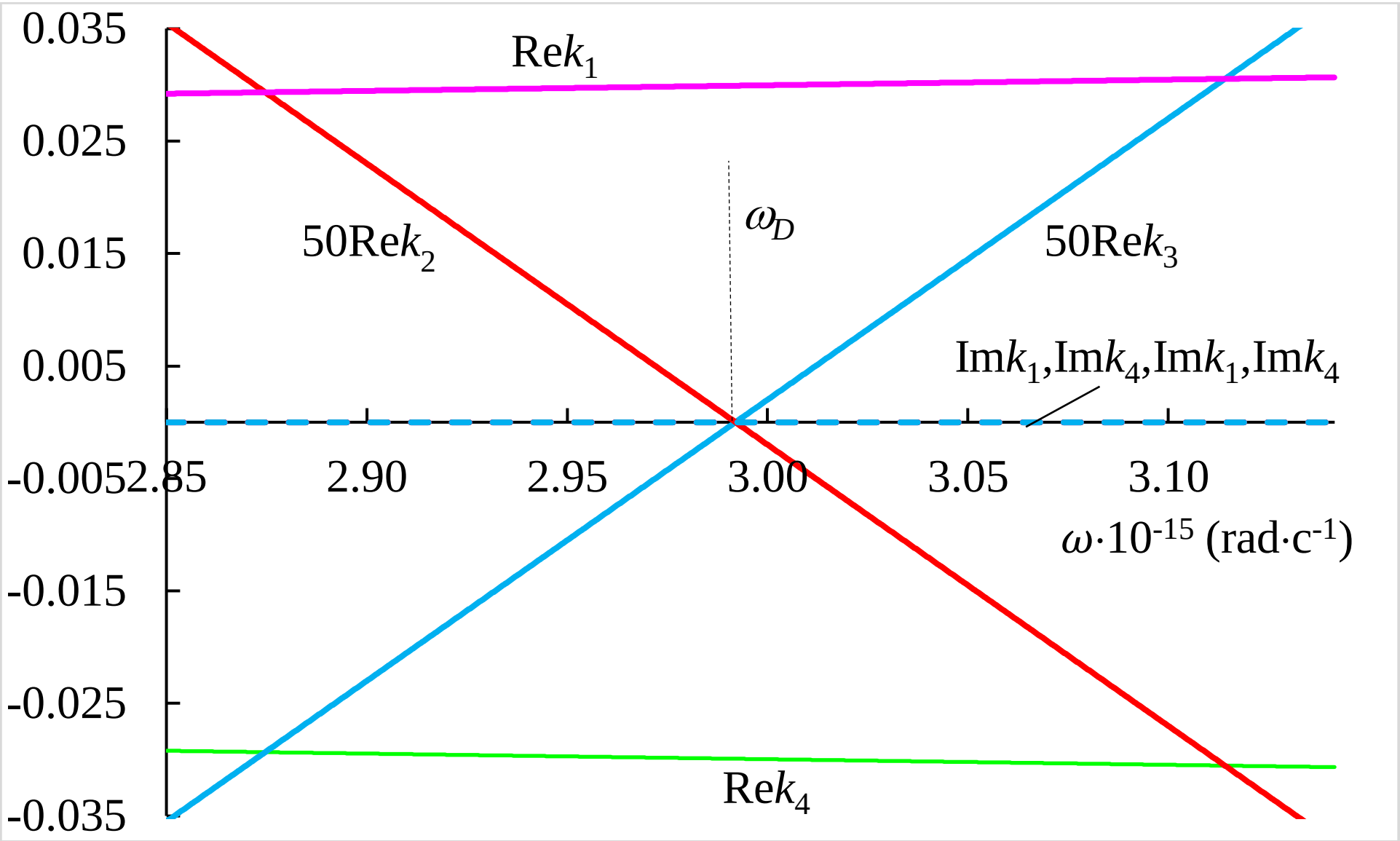

Fig.2. The dependences of $\mathrm{Re}k_{mz}$ and $\mathrm{Im}k_{mz}$ ($m$=1,2,3,4) for the frequency $\omega$. $\varepsilon = 2.25$, $p$=420 nm.

Now, equating $k_{1z}$ to $k_{2z}$ we obtain expression $\gamma = 0$. From this condition we obtain the new frequency, namely $\omega_3 = \frac{2ac\sqrt{|\varepsilon_m|}}{|\Delta|}$. This is the high frequency boundary of the new type PBG. The situation when $\gamma$ can stand fully imaginary and $\gamma = 0$ defined the boundary of this region is illustrated in Fig.3. In the region $\omega < \omega_3$ we have $\mathrm{Re}k_{1z} = \mathrm{Re}k_{2z} \neq 0$, with $\mathrm{Im}k_{1z} = \mathrm{Im}k_{2z}$, and $\mathrm{Re}k_{3z} = \mathrm{Re}k_{4z} \neq 0$, with $\mathrm{Im}k_{3z} = \mathrm{Im}k_{4z}$, that is here we have indirect PBG, unlike direct PBG located in region $\omega > \omega_1$ (see, also [24]). And, of course, the frequency $\omega_3$ is not the Dirac point frequency, too.
Now we consider the case when helically structured 1D PC being in external magnetic field directed along the helix axis and the medium has a magneto-optical activity, that is we consider the case when time-reversal symmetry is broken. In this case the dielectric permittivity and magnetic permeability tensors have the forms:

$$\hat{\varepsilon}(z) = \varepsilon_m \begin{pmatrix} 1 + \delta\cos 2az & \pm\delta\sin 2az \pm ig/\varepsilon_m & 0 \\ \pm\delta\sin 2az \mp ig/\varepsilon_m & 1 - \delta\cos 2az & 0 \\ 0 & 0 & 1 - \delta \end{pmatrix},\ \hat{\mu}(z) = \hat{I}, \quad (5)$$

where $g$ is the parameter of magneto-optical activity, it is the function of Verdet constant, external magnetic field, incident light wavelength and dielectric permittivity of media. And the dispersion equation has the form:

$$\left(\frac{\omega^2}{c^2}\varepsilon_1 - k_{mz}^2 - a^2\right)\left(\frac{\omega^2}{c^2}\varepsilon_2 - k_{mz}^2 - a^2\right) - \left(2ak_{mz} - \frac{\omega^2}{c^2}\mathrm{g}\right)^2 = 0. \quad (6)$$

In this case at $\varepsilon_1 = \varepsilon_2 = \varepsilon$ again, as above, Eq. (6) splits into two quadratic equations that have roots, already in the forms:

$$k_{1z} = \frac{\omega}{c}\sqrt{\varepsilon - \mathrm{g}} + a,\ \ k_{2z} = \frac{\omega}{c}\sqrt{\varepsilon + \mathrm{g}} - a, \qquad (7)$$
$$k_{3z} = -\frac{\omega}{c}\sqrt{\varepsilon - \mathrm{g}} + a,\ \ k_{4z} = -\frac{\omega}{c}\sqrt{\varepsilon + \mathrm{g}} - a.$$

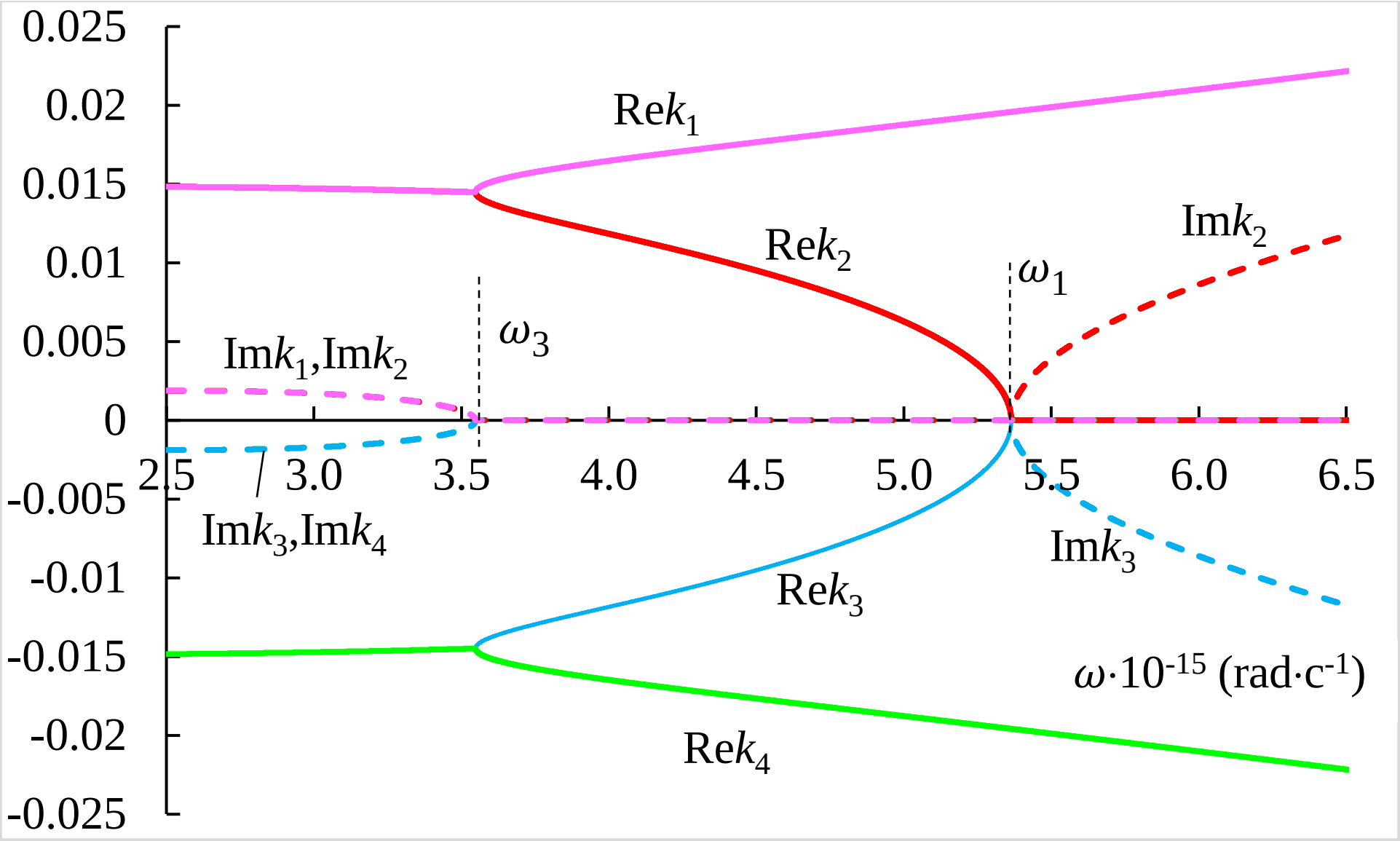

Fig.3. The dependences of Re$k_{mz}$ and Im$k_{mz}$ ($m$=1,2,3,4) for the frequency $\omega$. $\varepsilon_1 = -0.9.\, \varepsilon_2 = 0.7$, $p$=420 nm.

Fig. 4 shows the dependences of $k_{mz}$ ($m$=1,2,3,4) for the frequency $\omega$ in the region where intersects the curves $k_{2z}$ and $k_{3z}$. Thus, we have the intersection of the wave vector curves, again and the linear law of the wave vector dependence on the frequency, too, i.e., here we have a Dirac point. Equating $k_{2z}$ to $k_{3z}$ we obtain:

$$\frac{\omega}{c}\left(\sqrt{\varepsilon - \mathrm{g}} + \sqrt{\varepsilon + \mathrm{g}}\right) = 2a. \qquad (8)$$

From this equation for Dirac point frequency $\omega_{D1}$ we obtain (at the condition $\mathrm{g} \ll \varepsilon$)

$$\omega_{D1} = p\sqrt{\varepsilon}\left(1 + \frac{\mathrm{g}^2}{8\varepsilon^2}\right). \qquad (9)$$

We have obtained exactly the same equation as in [24] for the PBG frequency (which is simultaneously the Dirac point frequency in this case), but with the help of another assumption than that in [24].

But it is not the only Dirac point emerging in helically structured PCs in this case. So, now already from condition $k_{1z} = k_{2z}$ we obtain

$$\frac{\omega}{c}\left(\sqrt{\varepsilon + \mathrm{g}} - \sqrt{\varepsilon - \mathrm{g}}\right) = 2a. \qquad (10)$$

From this equation for new Dirac point frequency $\omega_{D2}$ we obtain (at the condition $\mathrm{g} \ll \varepsilon$)

$$\omega_{D2} = \frac{2ac}{\mathrm{g}}\sqrt{\varepsilon}. \qquad (11)$$

Fig. 5 shows the dependences of $k_{zm}$ ($m$=1,2,3,4) for the frequency $\omega$ around the frequency $\omega_{D2}$. At the presence of anisotropic absorption (Re$\varepsilon_1$=Re$\varepsilon_2$ but Im$\varepsilon_1 \neq$ Im$\varepsilon_2$) here was observed magnetically induced linear, nonreciprocal and tunable, transparency [26-28].

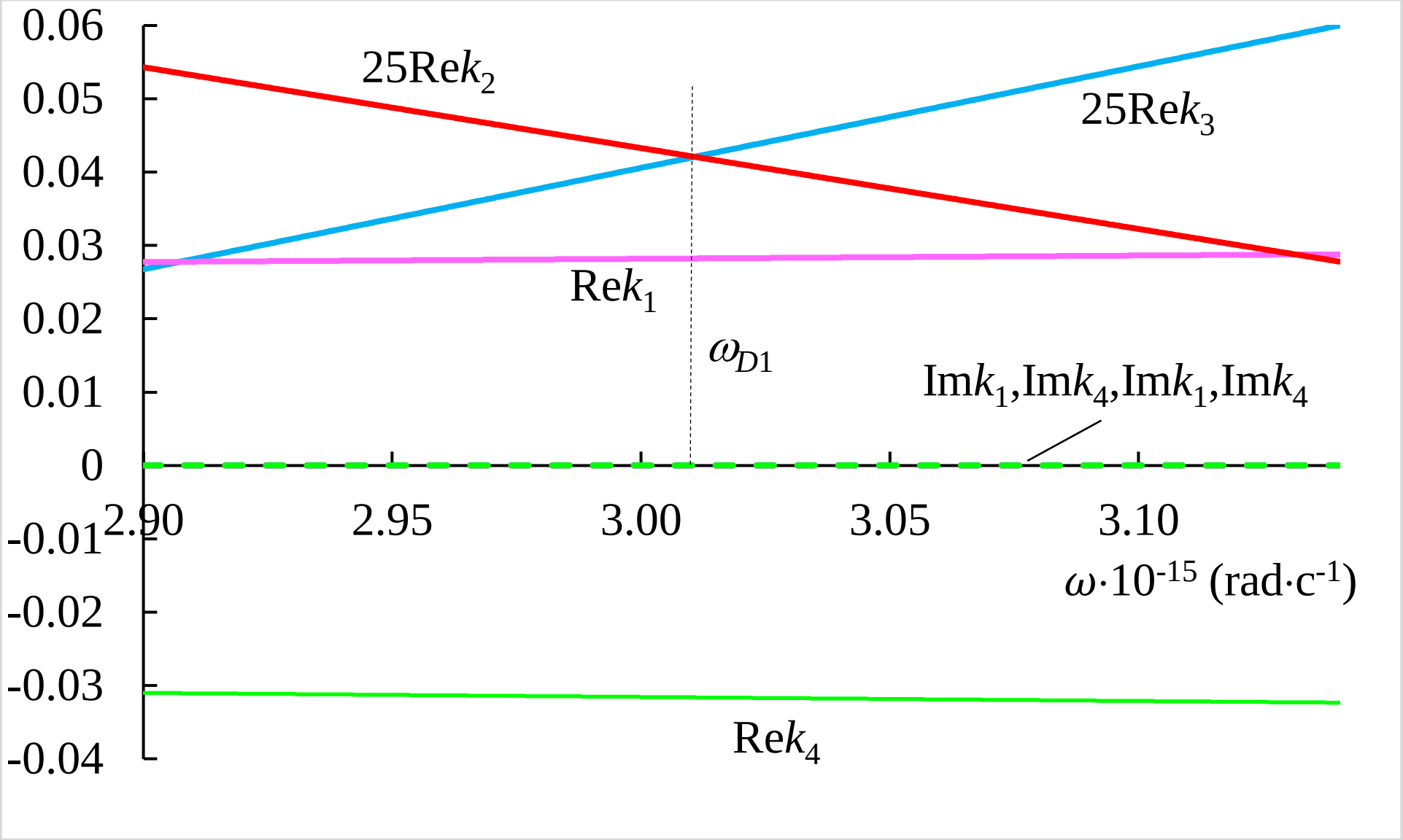

Fig.4. The dependences of Re$k_{mz}$ and Im$k_{mz}$ ($m$=1,2,3,4) for the frequency $\omega$. $\varepsilon = 2.25$, g=0.5.

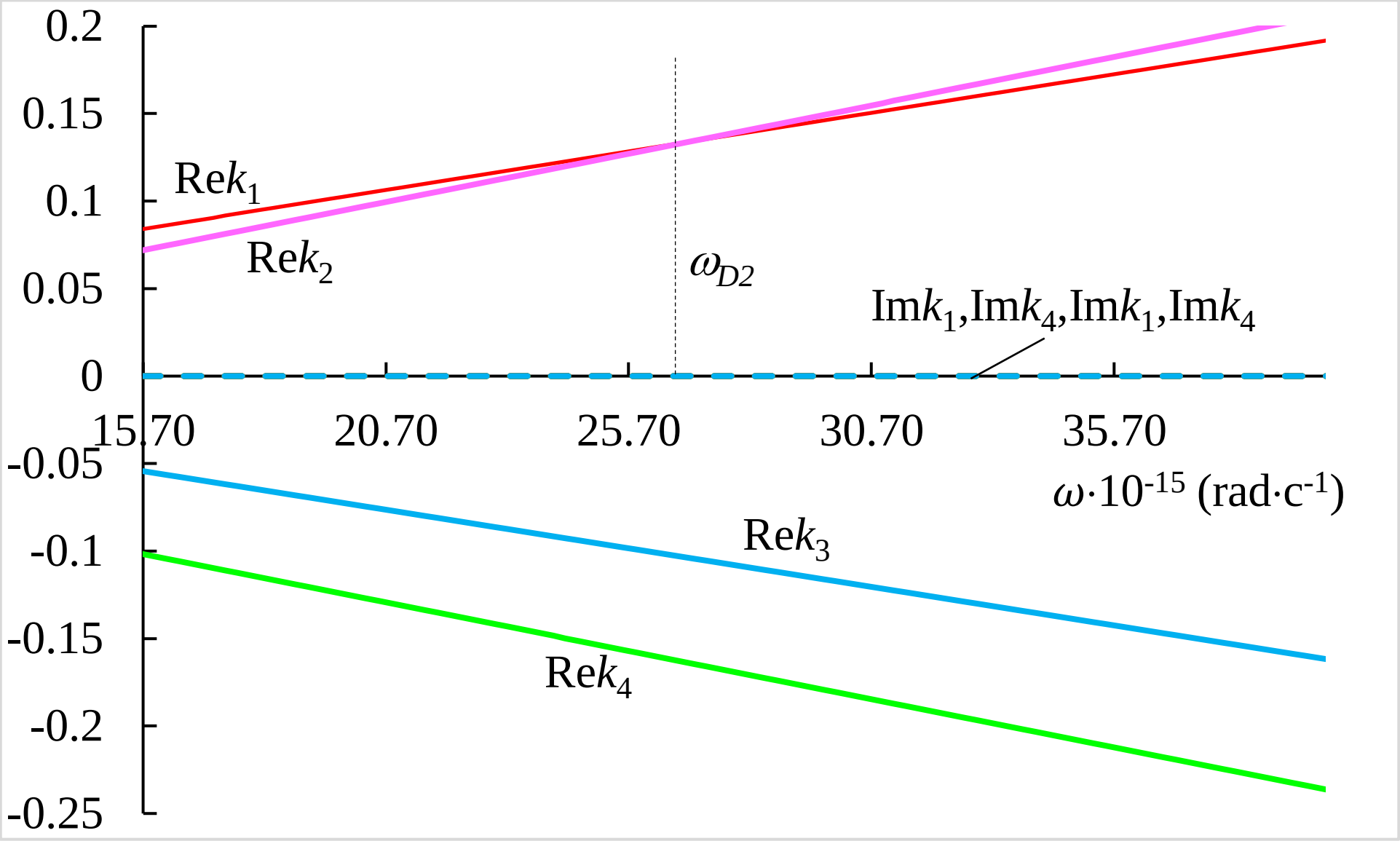

Fig.5. The dependences of Re$k_{mz}$ and Im$k_{mz}$ ($m$=1,2,3,4) for the frequency $\omega$. $\varepsilon = 2.25$, g=0.5.

But here one more interesting effect emerges too. Solving the boundary-value problem, at normal light incidence on the planar helically structured PC layer we can determine the reflection $R = {|E_r|^2}/{|E_i|^2}$, transmission $T = {|E_t|^2}/{|E_i|^2}$, and absorption $A = 1 - (R + T)$ coefficients.

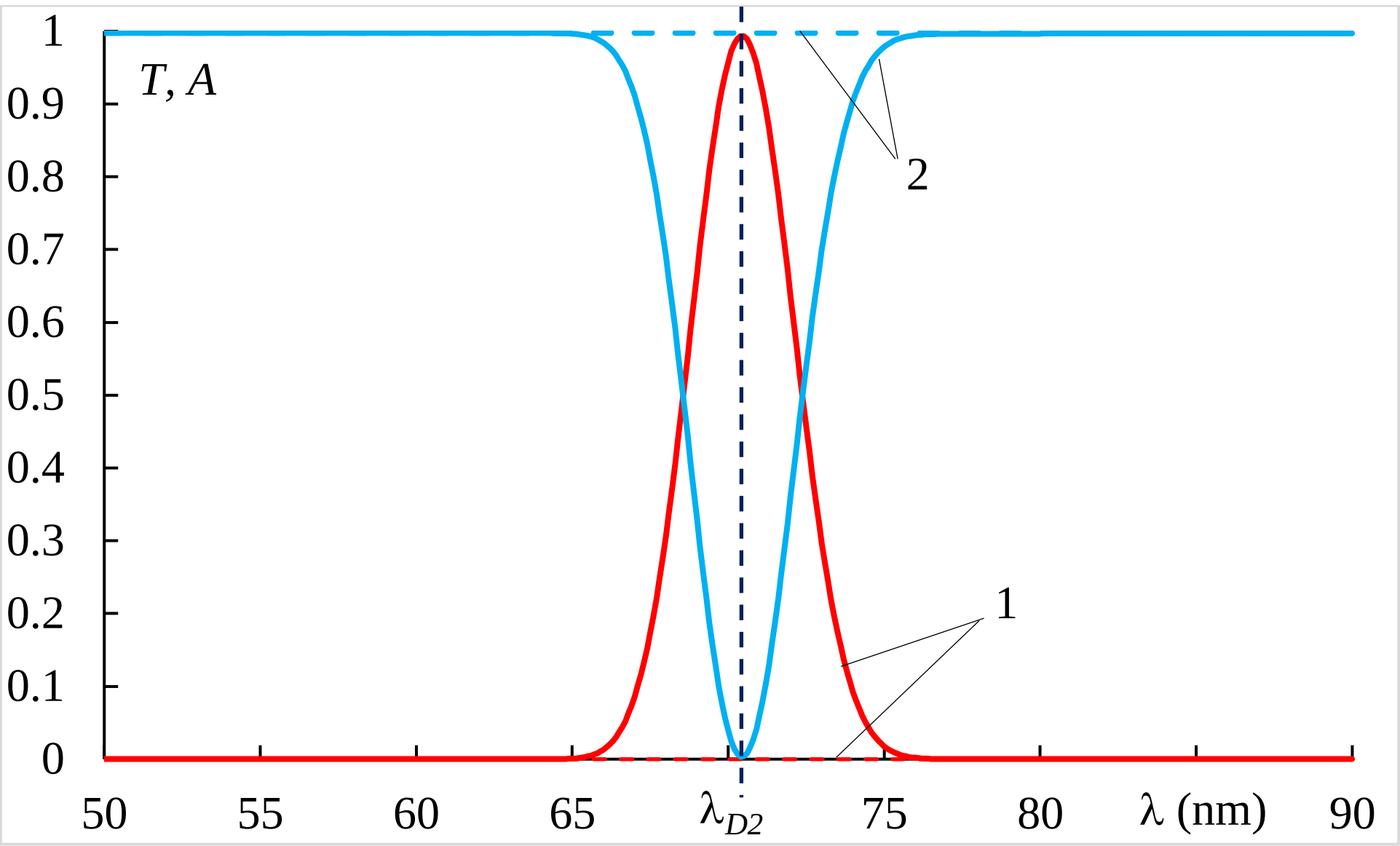

Fig.6. The spectra of $T$ transmission (curves 1) and $A$ absorption (curves 2. Solid lines correspond to the case g=0.2 and dashed lines to the case g= –0.2. $\varepsilon_1 = 2.29 + i0.1$; $\varepsilon_2 = 2.143 + i0.$; $p$=420 nm; Helically structured PCs layer thickness $d$=15$p$, PCs surroundings refractive index $n_s = \sqrt{\varepsilon}$.

Fig.6 shows the spectra of $T$ transmission (curves 1) and $A$ absorption (curves 2) in far short wavelength region around the Dirac point $\omega_{D2}$. Solid lines correspond to the case g=0.5 and dashed lines to the case g= –0.5. As it is seen in Fig. 6 the transmission spectrum of planar helically structured PC with locally anisotropic absorption (namely this case was considered by us) being in a longitudinal magnetic field in its shortwave part has a sharp peak with values of $T = 1$ on the top of this peak and correspondingly, absorption spectrum here has a deep dip with $A = 0$ at its bottom. Here we have $R$=0. At this wavelength, the medium becomes invisible.

In addition, it is seen in Fig. 6, on this wavelength we have $\Delta T = T_{forward} - T_{bacward} = -\Delta A = 1$, that is we have an ideal optical diode. The forward signal passes fully through the helically structured PC layer, while backward signal completely absorbs in this layer and there are not reflected wave, and moreover $\Delta R = 0$. And it takes please already starting from $d$=2$p$. Let us note, that, an optical diode of this type, in particular, can significantly reduce noise caused by reflections from numerous surfaces in transmission lines of optical communication systems.

In conclusion we showed theoretically the existence of a Dirac points in both helically structured 1D PCs and in magnetoactive helically structured 1D PCs. In addition, these points can find interesting applications and here observed interesting effects (unusual PBD, magnetically induced tunable transmission and absorption, ideal unidirectional transmission and etc).

## Acknowledgment

The work was supported by the Foundation for the Advancement of Theoretical Physics and Mathematics “BASIS” (Grant № 21-1-1-6-1).